\documentclass[prl,aps,showpacs,epsf,superscriptaddress,twocolumn,nobibnotes]{revtex4}
\usepackage{color}
\usepackage{amsmath}
\usepackage{amssymb}
\usepackage{graphicx}
\usepackage{bm}
\usepackage{hyperref}

\def\be{\begin{equation}}
\def\ee{\end{equation}}

\begin{document}

\title{Critical Velocity and Dissipation of an ultracold Bose-Fermi Counterflow}

\author{Marion Delehaye}\thanks{These two authors contributed equally to this work}
\author{S\'ebastien Laurent}\thanks{These two authors contributed equally to this work}
\author{Igor Ferrier-Barbut}
\altaffiliation{Present Address: 5. Physikalisches Institut and Center for Integrated Quantum Science and Technology,
Universit\"at Stuttgart, Pfaffenwaldring 57, 70550 Stuttgart, Germany}
\author{Shuwei Jin}
\author{Fr\'ed\'eric Chevy}
\author{Christophe Salomon}
\affiliation{Laboratoire Kastler Brossel, ENS-PSL, CNRS, UPMC-Sorbonne Universit\'es, and Coll\`ege de France, Paris, France}

\begin{abstract}
We study the dynamics of counterflowing  bosonic and fermionic lithium atoms. First, by tuning the interaction strength we measure the critical velocity $v_c$ of the  system in the BEC-BCS crossover in the low temperature regime and we compare it to the recent prediction of Castin {\em et al.}, Comptes Rendus Physique,
{\bf 16}, 241 (2015). Second, raising the temperature of the mixture slightly above the superfluid transitions reveals an unexpected phase-locking of the oscillations of the clouds induced by dissipation.

\end{abstract}


\maketitle
Superconductivity and superfluidity are spectacular macroscopic manifestations of quantum physics  at low temperature. Besides liquid helium 4 and helium 3, dilute quantum gases have emerged over the years as a versatile tool to probe superfluid properties in diverse and controlled situations. Frictionless flows have been observed with both bosonic and fermionic atomic species, in different geometries and in a large range of interaction parameters from the weakly interacting Bose gas to strongly correlated fermionic systems~\cite{raman1999evidence,Chikkatur2000Suppression, ryu2007observation, miller2007critical, desbuquois2012superfluid, weimer2014critical}. Several other hallmarks of superfluidity such as quantized vortices or second sound were also observed in cold atoms~\cite{madison2000vortex, zwierlein2005vortices, sidorenkov2013second}.

A peculiar feature of superfluid flows is the existence of a  critical velocity above which dissipation arises. In Landau's original argument, this velocity is associated with the threshold for creation of elementary excitations in the superfluid: for a linear dispersion relation, it predicts that the critical velocity is simply given by the sound velocity in the quantum liquid. This critical velocity has been measured both in superfluid helium~\cite{wilks1987helium} and ultracold atoms~\cite{raman1999evidence, miller2007critical, desbuquois2012superfluid, weimer2014critical, Hulet2010dissipative}. However the recent production of a  Bose-Fermi double superfluid~\cite{ferrier2014mixture} raised new questions on Bose-Fermi mixtures~\cite{ozawa2014chandrasekhar, zhang2014calibration, cui2014atom, kinnunen2015induced} and interrogations on the validity of Landau's argument in the case of  superfluid counterflow~\cite{ abad2014counter, castin2015landau,   zheng2014quasiparticle, wen2014structure,  chevy2015counterflow, huitao2015landau}.

 In this letter,  we study the dynamics of a Bose-Fermi superfluid counterflow in the BEC-BCS crossover and at finite temperature. We show how friction arises when the relative velocity of the Bose and Fermi clouds increases and we confirm that damping occurs only above a certain critical relative velocity $v_{\rm c}$. We compare our measurements  to Landau's prediction and its recent generalization  $v_{\rm c}=c_{\rm s}^{\rm F}+c_{\rm s}^{\rm B}$ where $c_{\rm s}^{\rm F}$  and  $c_{\rm s}^{\rm B}$ are the sound velocities of the fermionic and bosonic components respectively~\cite{castin2015landau}. Finally, we study finite temperature damping of the counterflow and we show that the system can be mapped onto a Caldeira-Leggett-like model~\cite{caldeira1983path} of two quantum harmonic oscillators  coupled to a bath of excitations. This problem has been recently studied  as a toy model for decoherence in quantum networks~\cite{chou08exact} or for heat transport in crystals~\cite{zurcher90quantum} and we show here that the emergence of dissipation between the two clouds leads to a Zeno-like effect which locks their relative motions.

Our Bose and Fermi double-superfluid set-up was previously described in~\cite{ferrier2014mixture}. We prepare vapors of bosonic ($\rm B$) $^7\rm{Li}$ atoms spin-polarized in the second-to-lowest energy state and fermionic ($\rm F$)  $^6\rm{Li}$ atoms prepared in a balanced mixture of the two lowest spin states noted $|\!\!\uparrow\rangle$, $|\!\!\downarrow\rangle$. The two species are kept in the same cigar-shaped hybrid magnetic-optical trap in which evaporative cooling  is performed in the vicinity of the $832$\,G $^6$Li Feshbach resonance ~\cite{zurn2013feshbach}. The final number of fermions $N_{\rm F}= 2.5 \times 10^5 $ greatly exceeds that of the bosons $N_{\rm B}\sim 2.5\times 10^4 $ and the temperature of the sample is adjusted by stopping the evaporation at different trap depths. The  thermal pedestal surrounding the $^7$Li Bose-Einstein Condensate (BEC) provides a convenient low temperature thermometer for both species after sufficiently long thermalization time ($\sim 1$\, second). The lowest temperature achieved in this study corresponds to almost entirely superfluid clouds with  $T/T_{\rm c,\alpha=B,F}\leq 0.5$, where $T_{\rm c,\alpha}$ is the superfluidity transition temperature of species $\alpha$.

 \begin{figure}
 \includegraphics[width=0.95\columnwidth]{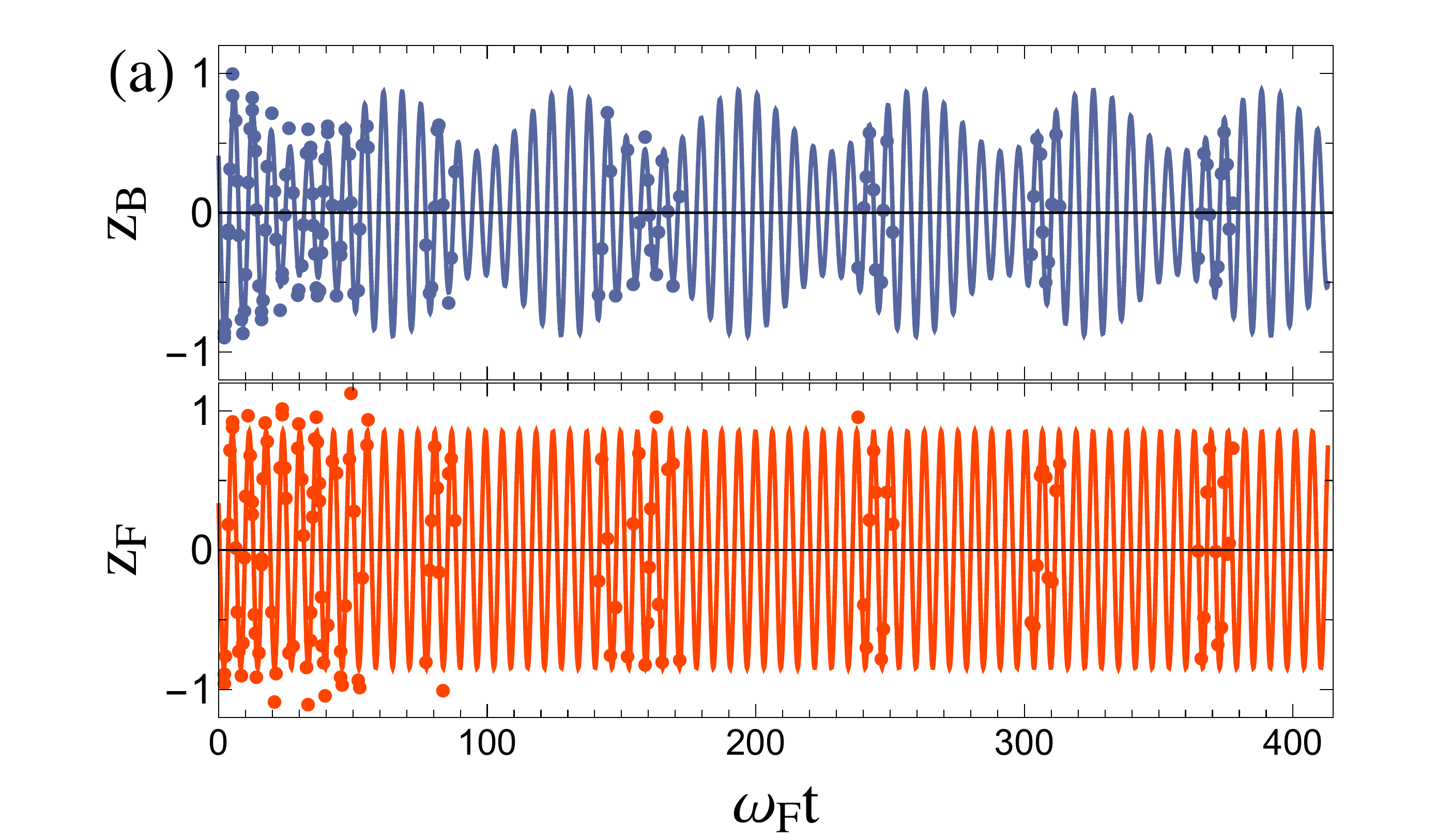}\\
 \includegraphics[width=0.95\columnwidth]{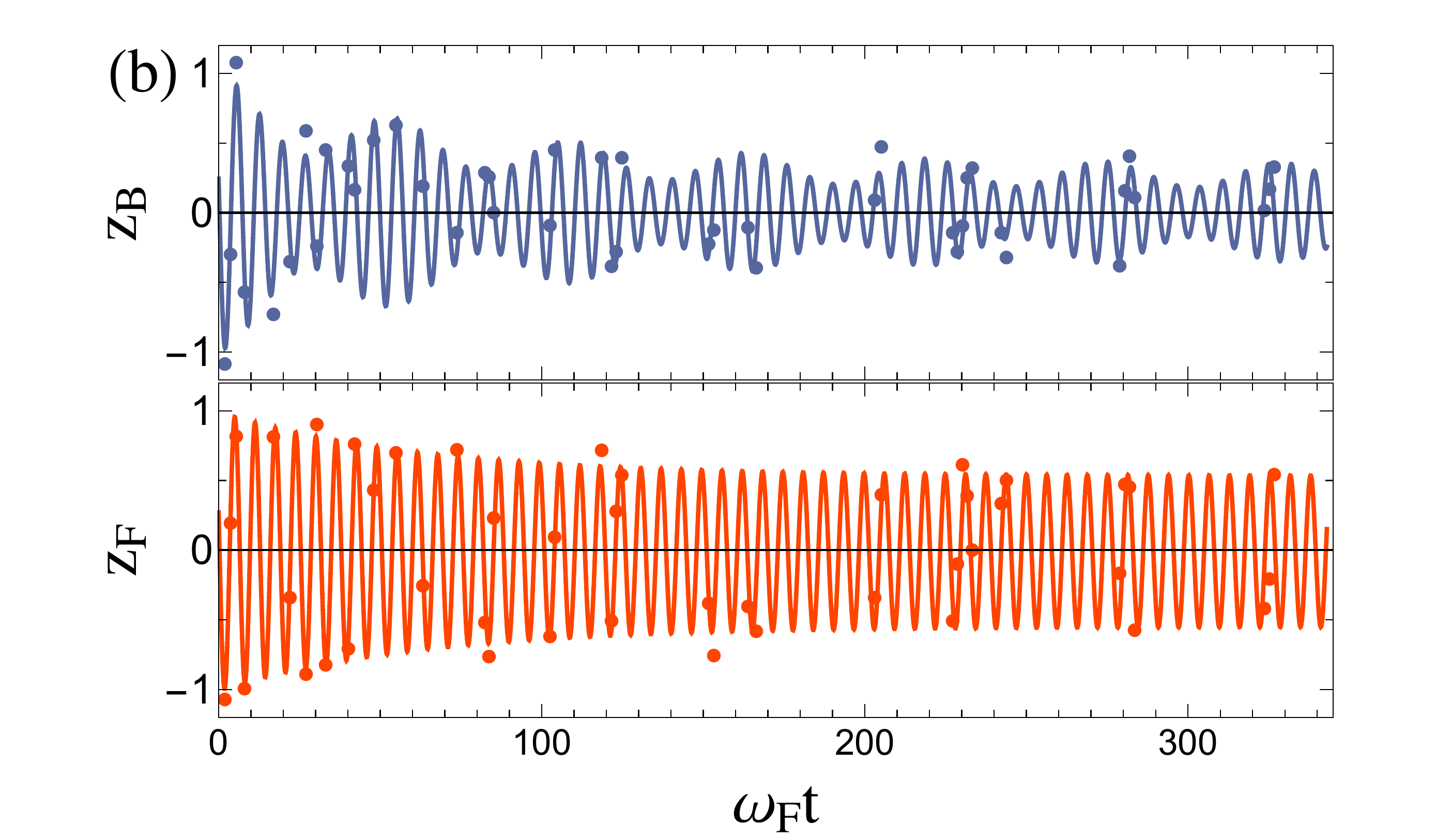}\\
 \includegraphics[width=0.95\columnwidth]{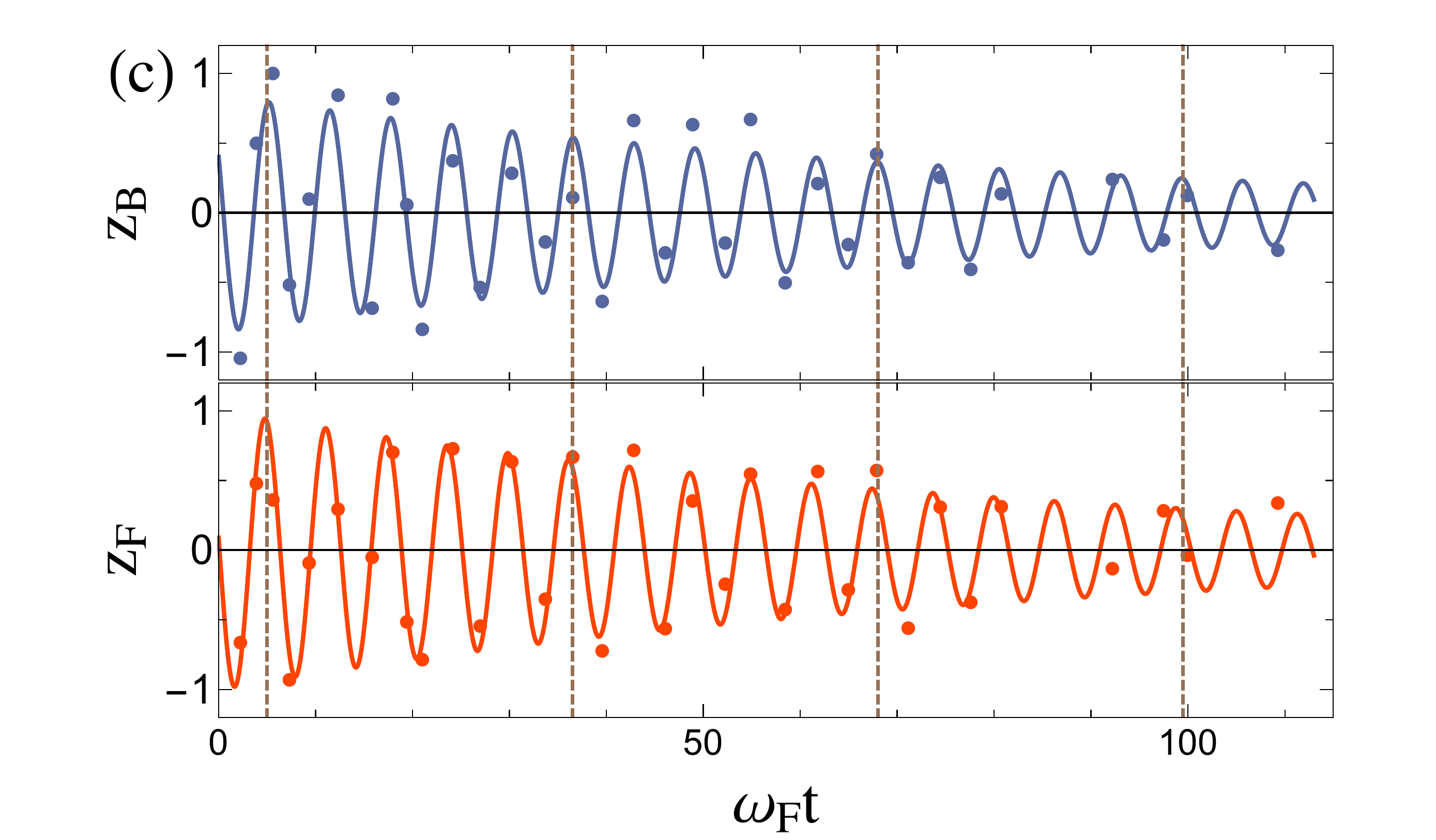}
 \caption{Center-of-mass oscillations of bosons (blue, top) and fermions (red, bottom), for different sets of parameters at unitarity. Solid lines: fits using Eq.\ref{eq:fitfunc} for the bosons and a similar equation for the fermions. (a) $T/T_{\rm F}=0.03$, $T/T_{c,b}\leq 0.5$, $z_0=10\,\rm{\mu m}$. Superfluid regime, no damping is observed and  $\omega_{\rm B}=2\pi\times 15.41(1)\,\rm{Hz}\approx \sqrt{6/7}\,\omega_{\rm F}$. The observed beating at $\omega_{\rm F}-\omega_{\rm B}$ is due to coherent energy exchange between the clouds.  (b) $T/T_{\rm F}=0.03$ and $z_0=150\,\rm{\mu m}$. For a larger initial displacement, initial damping ($\gamma_{\rm B}=2.4\,\rm{s^{-1}}$) is followed by steady-state evolution. $\omega_{\rm B}=2\pi\times 14.2(1)\,\rm{Hz}\approx \sqrt{6/7}\,\omega_{\rm F}$.  (c)  $T/T_{\rm F}=0.4$ and $z_0=80\,\rm{\mu m}$. At higher temperature, phase-locking of the two frequencies is observed with  $\omega_{\rm F} \approx \omega_{\rm B} =2\pi\times 17.9(3)\, \rm{Hz}$ and $\gamma_{\rm B}=\gamma_{\rm F}=1.4(5)\,\rm{s^{-1}}$.
 }
 \label{fig:fig1}
 \end{figure}

 The magnetic field values used in the experiment ($780-880$\,G) enable us to scan the fermion-fermion interaction within a range $-0.5 \leq 1/k_{\rm F}a_{\rm F} \leq 1 $. Here, $a_{\rm F}$ is the s-wave scattering length between  $|\!\!\uparrow\rangle $ and  $|\!\!\downarrow\rangle $ fermions   and the Fermi momentum $k_{\rm F}$ is defined by $\hbar^2 k_{\rm F}^2/2m_{\rm F}= \hbar \bar\omega (3 N_{\rm F})^{1/3} $ with $\bar \omega $ the geometric mean of the trap frequencies, and $N_{\rm F} $ the total number of fermions of mass $m_{\rm F}$.
 In our shallowest traps, typical trap frequencies for $^6\rm{Li}$ are  $\omega_x=\omega_y= 2\pi\times 550 $\,Hz and $\omega_z= 2\pi \times 17$\,Hz. Since the bosonic and fermionic isotopes experience the same trapping potentials, the oscillation frequencies  of the two species are within a ratio $\sqrt{6/7}\simeq 0.9$.

 We excite the dipole modes of the system by displacing adiabatically the centers of mass of the clouds from their initial position by a distance $z_0$ along the weakly confined $z$ direction, and abruptly releasing  them in the trap.  The two clouds evolve for a variable time $t$ before  \textit{in situ} absorption images perpendicular to $z$ direction are taken. The measurement of their doubly integrated density profiles gives access to axial positions and atom numbers of both species. Typical time evolutions of the centers of mass are shown in Fig.\ref{fig:fig1} for different parameter values. Since the Bose and Fermi components oscillate at different frequencies,  they oscillate in quadrature after a few periods. By changing $z_0$, we can thus tune the maximum relative velocity between the two clouds and probe the critical superfluid counterflow.

 As shown in Fig.\ref{fig:fig1} a, the superfluid counterflow exhibits no visible damping on a $\simeq 5$\,s time scale for very low temperature and small initial displacement. A striking feature is the beat note on the $^7$Li oscillation amplitude due to the coherent mean-field coupling to the $^6$Li cloud~\cite{ferrier2014mixture}. For larger relative velocities, $^7$Li oscillations are initially damped (Fig.\ref{fig:fig1} b) until a steady-state regime as in Fig.\ref{fig:fig1} a is reached. We fit the time evolution of the cloud position using the phenomenological law

 \begin{align}
 \label{eq:fitfunc}
 z_{\rm B}(t)&=d(t)\left[a\cos(\omega_{\rm B}t)+b\cos(\omega_{\rm F}t)\right],\\
 d(t)&=d_1+d_2 \exp(-\gamma_{\rm B}t).\notag
 \end{align}

\begin{figure}
 \centering
 \includegraphics[width=\columnwidth]{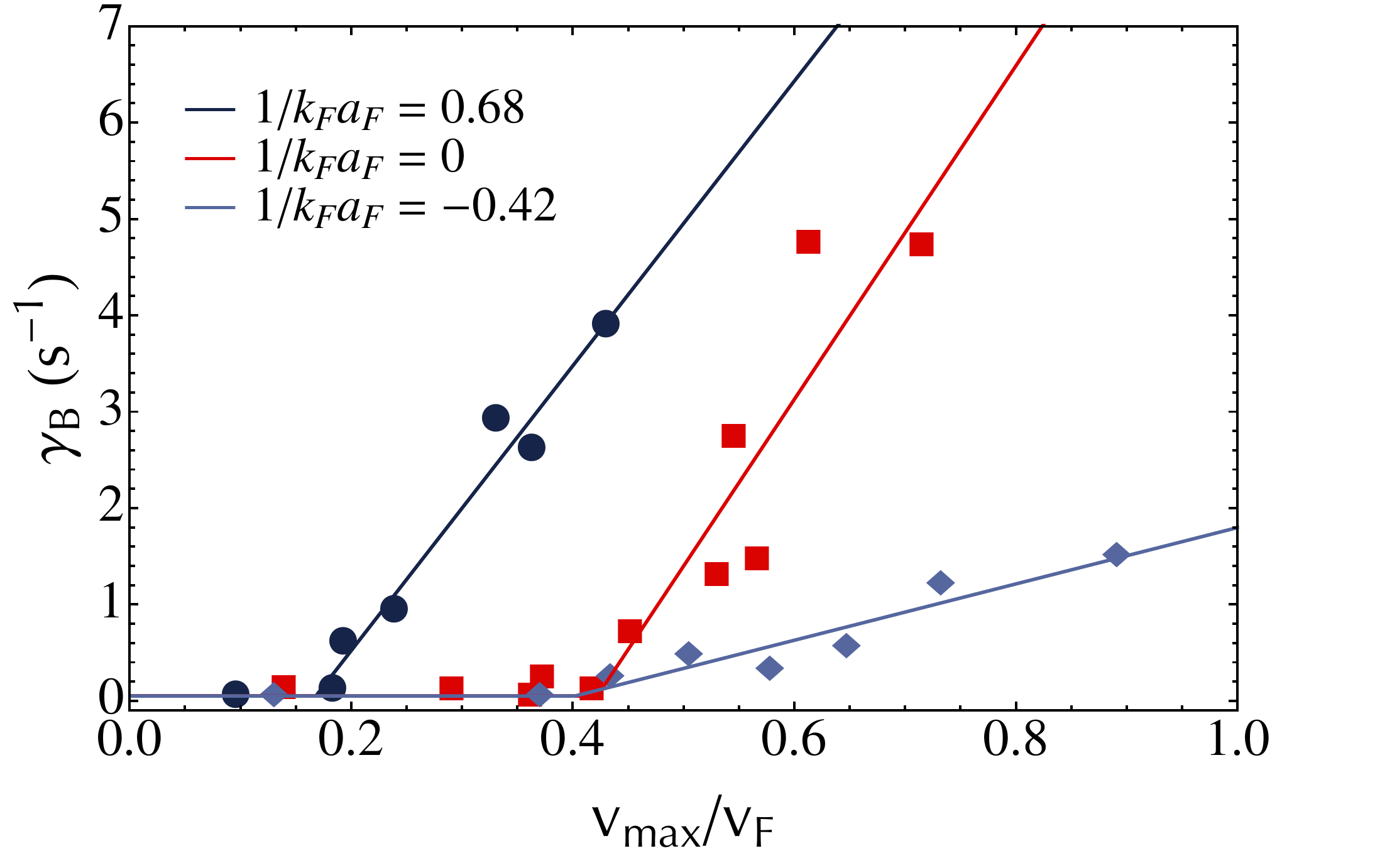}
 \caption{Damping rate of the center of mass oscillations versus maximal relative velocity in the BEC-BCS crossover in unit of the  Fermi velocity $v_{\rm F}$. Dark blue dots: BEC side (780\,G) $1/k_{\rm F}a_{\rm F}=0.68$, red squares: unitarity (832.2\,G) $1/k_{\rm F}a_{\rm F}=0$, light blue diamonds: BCS side (880\,G) $1/k_{\rm F}a_{\rm F}=-0.42$.  Power law fits with thresholds provide the critical velocity (solid lines). }
 \label{fig:fig4}
 \end{figure}

We measure the damping rate $\gamma_{\rm B}$ as a function of relative velocity for six different values of magnetic field, exploring a large region of the crossover going from the BCS  ($1/k_{\rm F}a_{\rm F}=-0.42$, $B=880\,$G) to the BEC side ($1/k_{\rm F}a_{\rm F}=0.68$, $B=780\,$G), see Fig.\ref{fig:fig4}. For these magnetic field values, the Bose gas remains in the weakly  interacting (repulsive) regime and the Bose-Fermi scattering length is $a_{\rm BF}\simeq 41a_0$, constant in this magnetic field range, and equal for both $|\!\!\uparrow\rangle$ and $|\!\!\downarrow\rangle$ spin states.

  \begin{figure}
 \centering
  \includegraphics[width=\columnwidth]{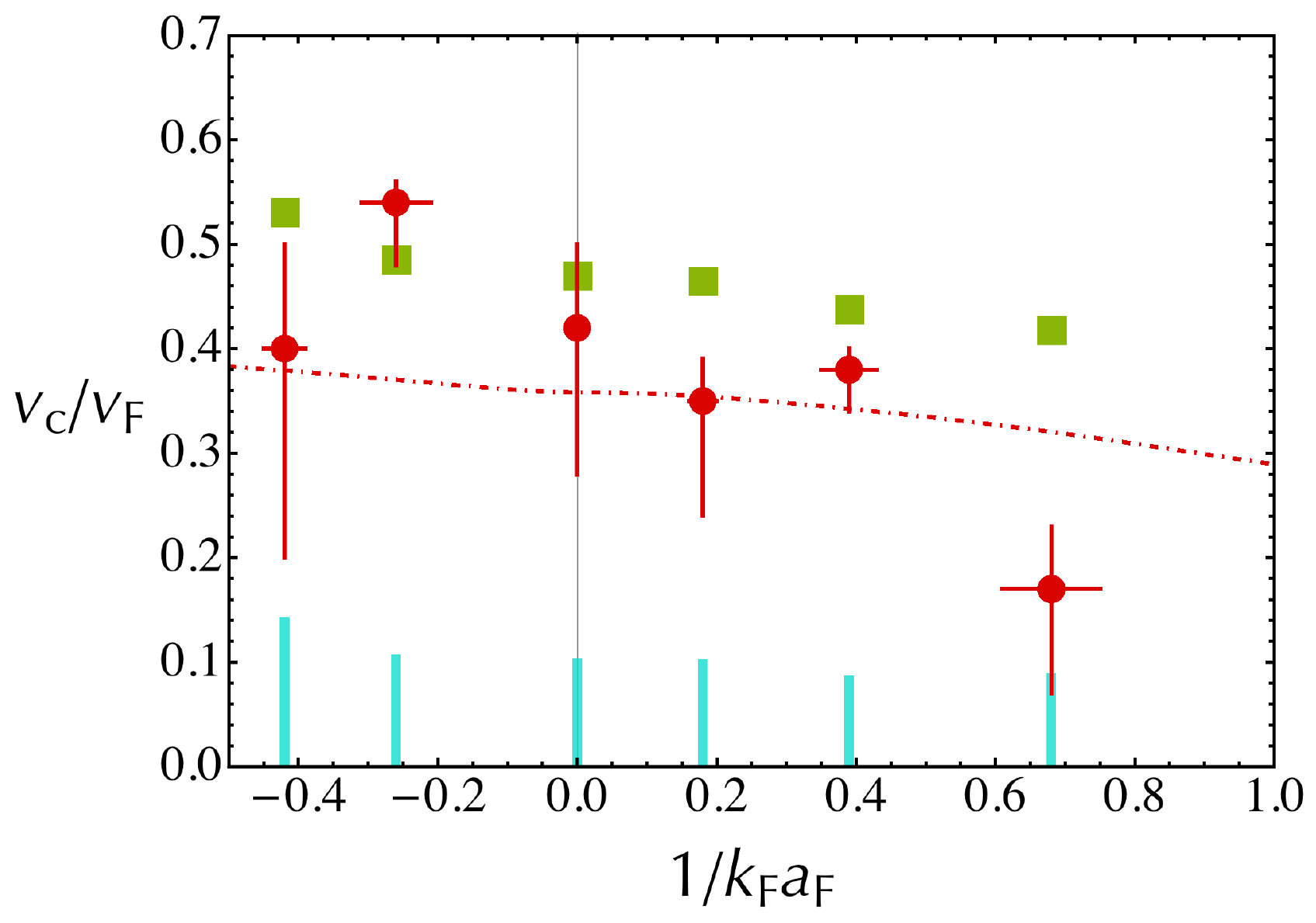}
   \caption{Critical velocity of the Bose-Fermi superfluid counterflow in the BEC-BCS crossover normalized to the Fermi velocity $v_{\rm F}$. Red dots: measurements. Red dot-dashed line: sound velocity $c^{\rm F}_s$ of an elongated homogeneous Fermi superfluid calculated from its equation of state~\cite{navon2010Ground,astrakharchik2014quantum} after integration of the density in the transverse plane, and also measured in~\cite{Thomas07soundvel}. Blue bars: calculated sound velocity $c^{\rm B}_s $ of the elongated $^7$Li BEC for each magnetic field, (880\,G, 860\,G, 832\,G, 816\,G, 800\,G, 780\,G). Green squares indicate the prediction $v_c=c^{\rm F}_s+c^{\rm B}_s $.   Error bars and $c^{\rm B}_s$ are discussed in \cite{suppl}.}
   \label{fig:fig5}
 \end{figure}

We extract the critical velocity $v_{\rm c}$  using an {\em ad-hoc} power-law fitting function $\gamma_{\rm B} = A \Theta (v-v_c) ((v-v_c)/v_{\rm F})^{\alpha}$, where $\Theta$ is the Heaviside function and $v_{\rm F}$ is the Fermi velocity given by $v_{\rm F}=\hbar k_{\rm F}/m_{\rm F}$. For details see \cite{suppl}. $v_{\rm c}$ in the BEC-BCS crossover is displayed in Fig.\ref{fig:fig5} (red dots) and  compared to Landau's and Castin {\em et al.} \cite{castin2015landau} predictions.  In this latter work, dissipation arises  by creation of excitation pairs and yields a critical velocity  $v_{\rm c}=\underset{\underset{\; \sigma\,=\,{\rm f, b}}{\boldsymbol{p}}}{{\rm Min}}\,\left(\frac{\epsilon^{\rm B}(\boldsymbol p)+\epsilon^{\rm F}_\sigma(\boldsymbol p)}{p}\right)$. In this expression,  $\epsilon^{\rm B}(\boldsymbol p)$ denotes the dispersion relation of excitations in the BEC and  $\epsilon^{\rm F}_\sigma(\boldsymbol p)$ refers to the two possible branches of the Fermi superfluid, phonon-like ($\sigma={\rm b}$) and threshold for pair breaking excitations ($\sigma={\rm f}$)~\cite{combescot2006collective}. For homogeneous gases, at unitarity and on the BEC side of the crossover, this critical relative velocity turns out to be simply the sum of the respective sound velocities of the Bose and Fermi superfluids, $v_{\rm c}=c_{\rm s}^{\rm F}+c_{\rm s}^{\rm B}$.   We thus plot in Fig.\ref{fig:fig5} the calculated sound velocities of both superfluids in an elongated geometry obtained by integration over the transverse direction~\cite{stringari1998dynamics,Capuzzi2006Sound,luo2007measurement, astrakharchik2014quantum,navon2010Ground} (red dashed line $c_{\rm s}^{\rm F}$, blue bars $c_{\rm s}^{\rm B}$). 
 Typically $c_{\rm s}^{\rm B}$ contributes $\simeq 20-25\%$ to the sum shown as green squares in Fig.\ref{fig:fig5}. Around unitarity and on the BCS side of the resonance, our experimental data is consistent with this interpretation as well as with a critical velocity $v_c=c_s^{\rm F}$ that one would expect by considering the BEC as a single impurity moving inside the fermionic superfluid. By contrast, we clearly exclude the bosonic sound velocity as a threshold for dissipation.

Our measured critical velocities are significantly higher than those previously reported in pure fermionic systems which, for all interaction strengths, were lower than Landau's criterion~\cite{miller2007critical, weimer2014critical}.
 The main difference with our study is the use of focused laser beams instead of a BEC as a moving obstacle. In~\cite{weimer2014critical}, the laser beam  is piercing the whole cloud including its non-superfluid part where the density is low, and its potential may create a strong density modulation of the superfluid. These effects make a direct comparison to Landau criterion difficult~\cite{PalSingh15criticalvel}. On the contrary, in our system the size of the BEC (Thomas Fermi radii of $73,\,3,\,3\,\mu$m) is much smaller than the typical size of the Fermi cloud  ($350,\,13,\,13\,\mu$m around unitarity). For
  oscillation amplitudes up to  $\pm 200\, \rm{\mu m}$  the BEC probes only the superfluid core of the fermionic cloud.  During its oscillatory motion along $z$ the Bose gas may explore the edges of the Fermi superfluid where the density is smaller. However  it is easy to check  that the ratio $v/c_{\rm s}^{\rm F}$ is maximum when the centers of the two clouds coincide~\cite{suppl}. Finally, as the mean-field interaction between the two clouds is very small \cite{suppl} our BEC  acts as a  weakly interacting local probe of the Fermi superfluid.

 On the BEC side of the resonance (780\,G) however, we observe a strong reduction of the measured critical velocity compared to the predicted values. The effect is strikingly seen in Fig.\ref{fig:fig4}, dark blue dots (see also supplementary material~\cite{suppl}).
 This anomalously small value for positive scattering lengths is consistent with previous measurements~\cite{miller2007critical, weimer2014critical}. Its origin is still unclear but several explanations can be put forward~\cite{PalSingh15criticalvel}.
  Firstly, it is well known that vortex shedding can strongly reduce superfluid critical velocity. However, this mechanism requires a strong perturbation. The density of the Bose gas and the mean-field interaction between the two clouds are probably too small for vortex generation through a collective nucleation process. Second, inelastic losses increase on the BEC side of a fermionic Feshbach resonance and heat up the system \cite{regal2004lifetime}. This hypothesis is supported by the presence of a clearly visible pedestal in the density profiles of the BEC taken at 780\,G. At this value of the magnetic field, we measure a $\simeq 60$\% condensed fraction, corresponding to a temperature $T/T_{\rm c,B}\simeq 0.5$. Even though the two clouds are still superfluids as demonstrated by the critical behaviour around $v_c$, the increased temperature could be responsible for the decrease of $v_{\rm c}$.

 \begin{figure}
 \centering
 \includegraphics[width=\columnwidth]{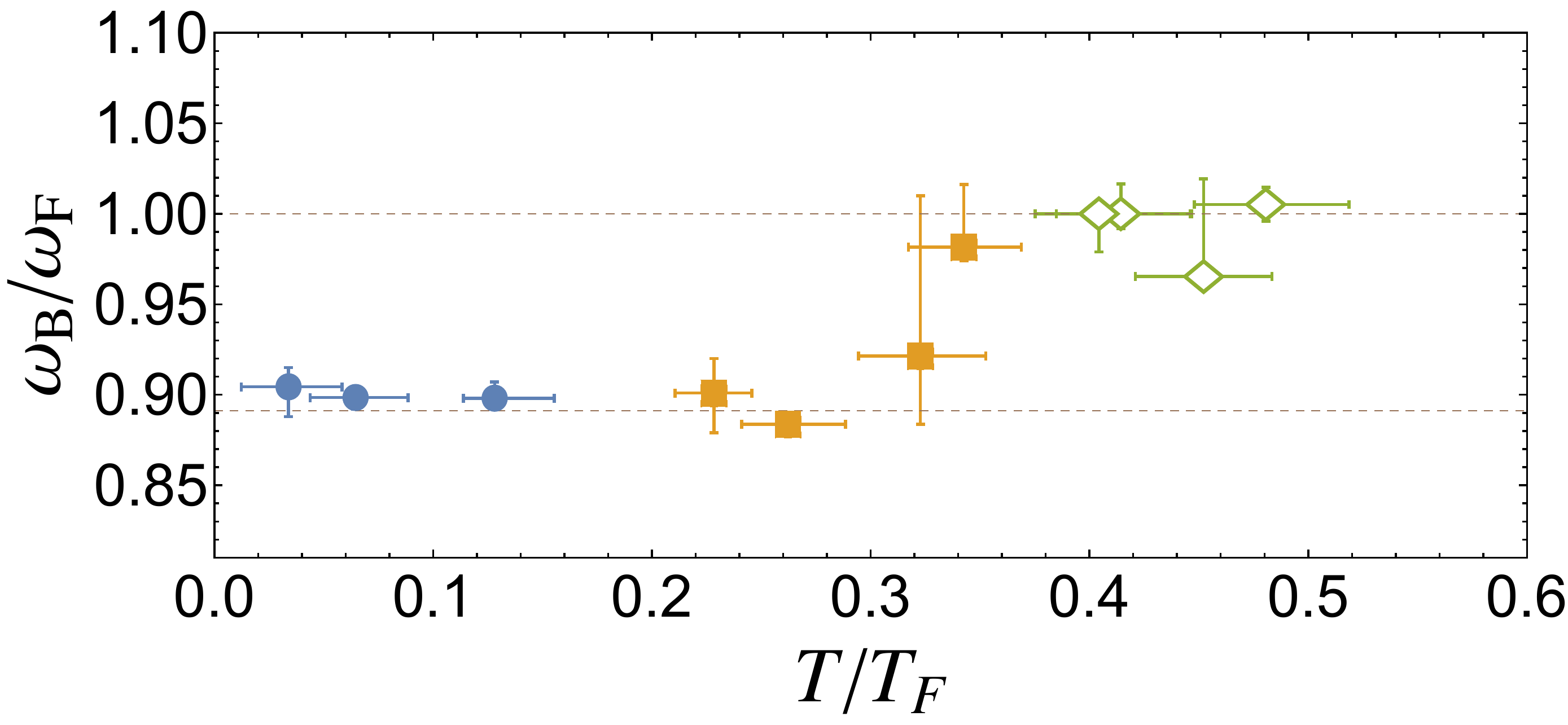}
 \caption{Ratio $\omega_{\rm B}/\omega_{\rm F}$ versus temperature of the cloud. Blue circles: the two clouds are superfluids. Yellow squares: only the bosonic component is superfluid. Green open diamonds: the two components are normal.   Above $T\approx  T_{\rm c,B}\approx 0.34\,T_{\rm F}>T_{\rm c,F}$, oscillations of the Bose and Fermi clouds become locked together at $\omega_{\rm F}$. Oscillations frequencies are obtained using a Lomb-Scargle algorithm~\cite{suppl}. The lower dashed line is the prediction of a low temperature mean field model~\cite{ferrier2014mixture}.}
   \label{fig:fig2}
 \end{figure}

We now present results of experiments performed at higher temperature ($0.03\lesssim T/T_{\rm F}\lesssim 0.5$) for $B=835\,$G. For low temperatures ($T/T_{\rm F}\leq 0.2$),  the  two clouds remain weakly coupled and, as observed in Fig.\ref{fig:fig2}, the bosonic and fermionic components oscillate at frequencies in the expected ratio $\simeq 0.9\simeq \sqrt{6/7}$. A new feature emerges for $T\gtrsim  T_{\rm c,B}\approx 0.34\,T_{\rm F}>T_{\rm c,F}$ where both gases are in the normal phase. In this ``high" temperature regime,   the two clouds  are locked in phase: $^7$Li oscillates at $^6$Li frequency (Fig.\ref{fig:fig2}) and the two components are equally damped (Fig.\ref{fig:fig1}.c). This remarkable  behavior can be understood as a Zeno effect arising from the increased dissipation between the two components. Indeed  the system can be described  as a set of two harmonic oscillators describing respectively the macroscopic motion of the global center of mass of the system (Kohn's mode~\cite{kohn1961cyclotron}) and the relative motion of the two clouds~\cite{suppl}. These two degrees of freedom are themselves coupled to the ``bath" of the internal excitations of the two clouds (breathing mode, quadrupole modes, pair breaking excitations...).

\begin{figure}
\centerline{\includegraphics[width=0.8\columnwidth]{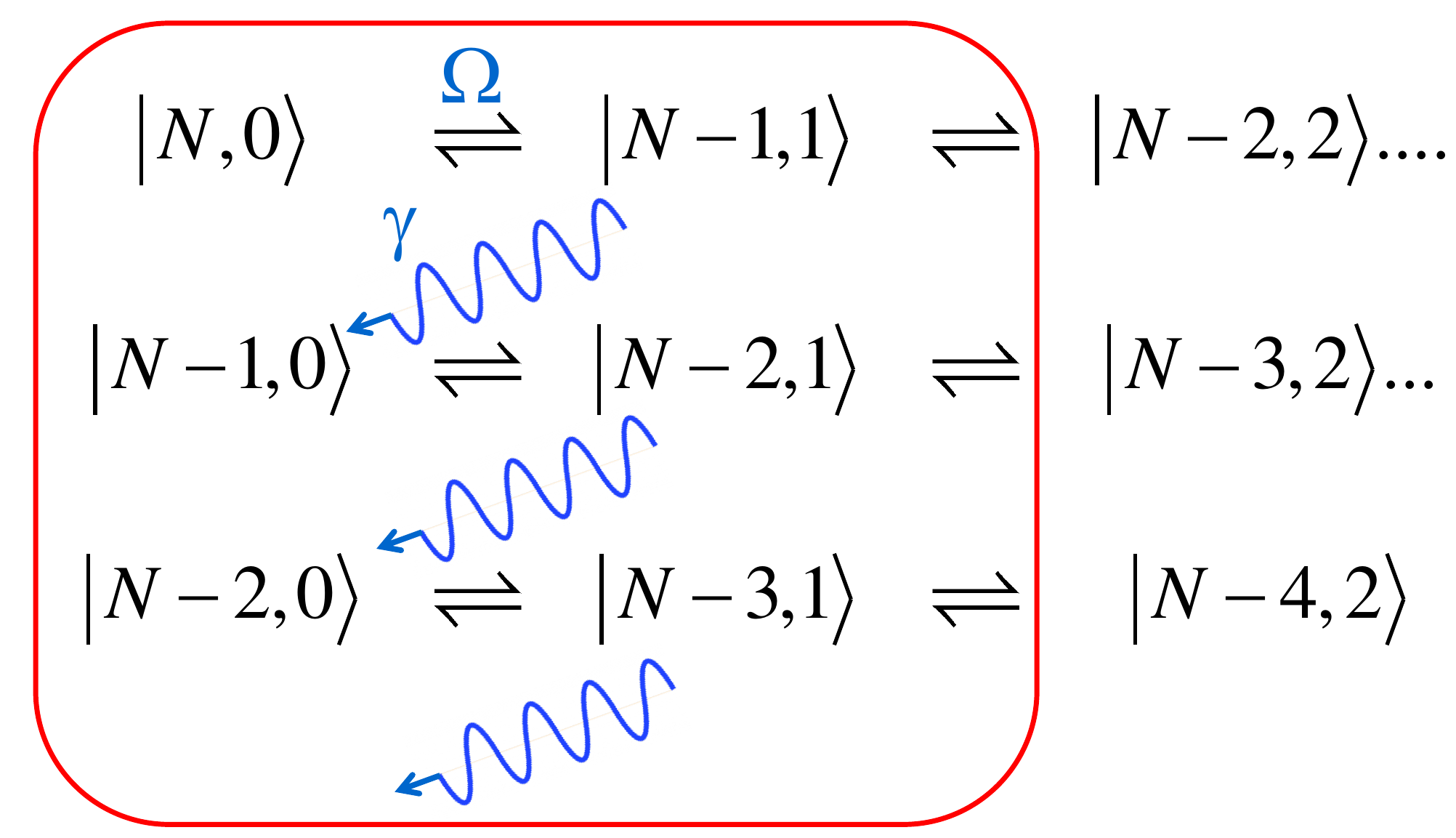}}
\caption{Radiative cascade of the center of mass motion. In $|N,n\rangle$, $N$ (resp. $n$) refers to the center of mass (resp. relative) motion of the two clouds (see text). When the decay rate of the relative motion is larger than the oscillation frequency difference between the two species, the dynamics is restricted to the center of mass degree of freedom: in this Zeno-like process, dissipation prevents excitation of the relative motion and the center of mass modes of the Bose and Fermi gases do not dephase.}
\label{fig:fig6}
\end{figure}

In the spirit of the dressed-atom picture, we can represent the state of the two harmonic oscillators by the ``radiative" cascade of Fig.\ref{fig:fig6}. Here the states $|N,n\rangle$ are labeled by the quantum numbers associated to Kohn's mode ($N$) and relative motion ($n$) of the two clouds and we trace-out the degrees of freedom of the bath. On the one hand,  Kohn's mode is not an eigenstate of the system for fermions and bosons of different masses; center-of-mass and relative-motion modes are coupled and this coherent coupling is responsible for the dephasing of the oscillations of the two clouds in the weakly interacting regime.  On the other hand interspecies interactions do not act on the center of mass of the whole system,  owing to Kohn's theorem, but on the contrary  lead to an irreversible ``radiative" decay of the {\it relative} motion at a rate $\gamma$.

In our experiments, the initial state is a pure center of mass excitation $|N,0\rangle$. If we neglect the interspecies coupling, the system evolves in the subspace spanned by $|N-n,n\rangle_{n=0,..,N}$ of the two coupled oscillators and the system oscillates at a frequency $\delta\omega\simeq \omega_{\rm B}-\omega_{\rm F}$ as the centers of mass of the Bose and Fermi clouds dephase. If we now consider the opposite limit where the decay rate $\gamma$ is larger than the dephasing frequency $\delta\omega$, the strong coupling to the bath prevents the conversion of the center of mass excitations into relative motion. As soon as the system is transferred into $|N-1,1\rangle$ it decays towards state $|N-1,0\rangle$. Similarly to optical pumping in quantum optics, we can eliminate adiabatically the excited states of the relative motion and restrict the dynamics of the system to the subspace $|N,0\rangle_{N=0..\infty}$ of Kohn's excitations. This situation is reminiscent of the synchronization of two spins immersed in a thermal bath predicted in \cite{Orth10Dynamics} or to phenomenological classical two-coupled oscillators model.

In this letter we have investigated how a Bose-Fermi superfluid flow is destabilized by temperature or relative velocity between the two clouds. In the limit of very low temperature the measured critical velocity for superfluid counterflow slightly exceeds the speed of sound of the elongated Fermi superfluid and decreases sharply towards the BEC side of the BEC-BCS crossover. In a future study,
 we will investigate the role of temperature, of the confining potential, and of the accelerated motion of the two clouds~\cite{PalSingh15criticalvel} that should provide a more accurate model for the damping rate versus velocity and more insights on the nature of the excitations. In particular the {\it ab initio} calculation of the damping rate will require to clarify the dissipation mechanism at play in a trapped system where the bandwidth of the excitation spectrum is narrow, in contrast to a genuine Caldeira-Leggett model~\cite{Onofrio2015Effective}.

\acknowledgments
The authors acknowledge support from R\'egion Ile de France (Atomix Project), ERC (ThermoDynaMix Project) and Institut de France (Louis D. Prize). They thank I. Danaila, N. Proukakis, K.L. Lee and M. Pierce for insightful comments and discussions, and J. Dalibard, Y. Castin, S. Nascimb\`ene and T. Yefsah for critical reading of the manuscript.

\bibliography{bibliographie}

\newpage

\onecolumngrid
\newpage

\centerline{\bf\sc Supplemental Information}

\bigskip

\section{Damping rates in the BEC-BCS crossover}
\begin{figure*}[h]
 \centering
 \includegraphics[scale=0.38]{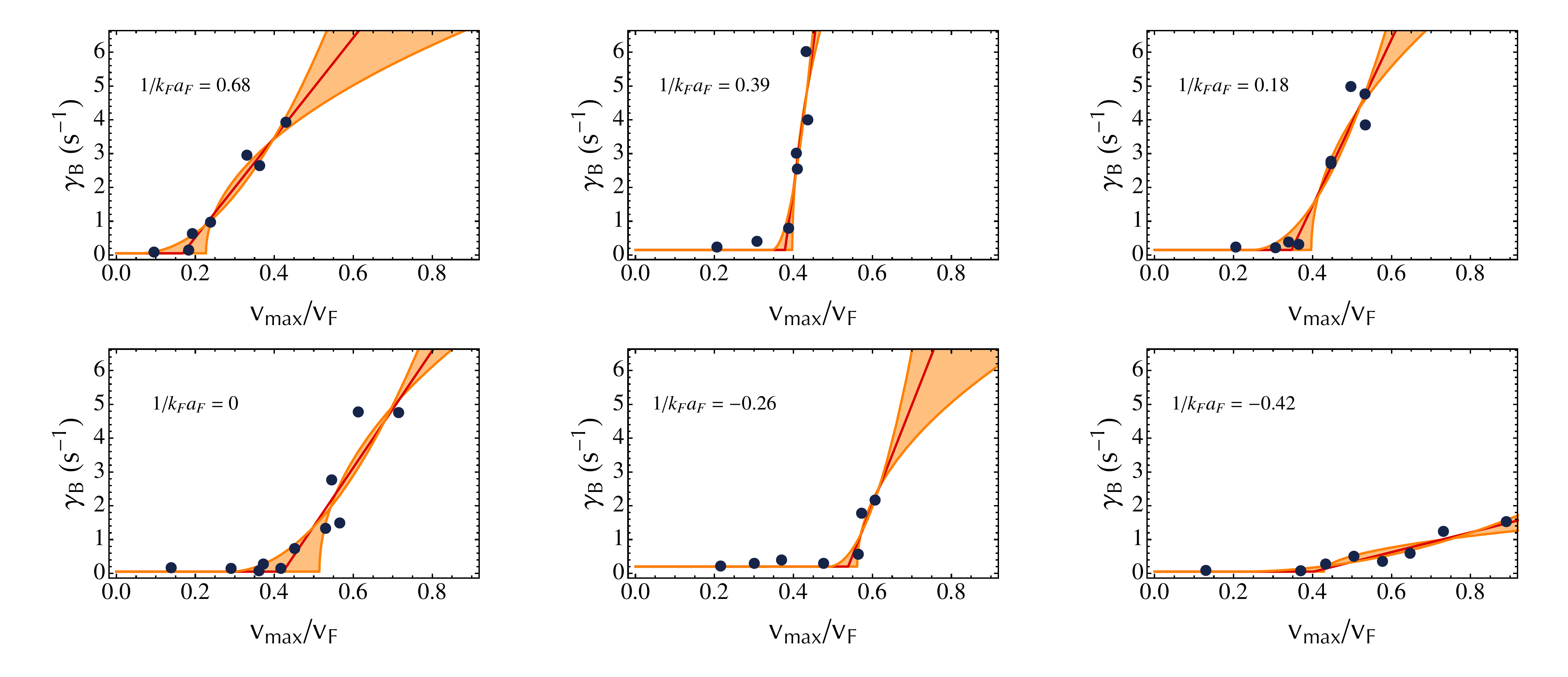}
 \caption{Damping rates of the center of mass oscillations versus maximal relative velocity in the BEC-BCS crossover in unit of the Fermi velocity $v_{\rm{F}}$. Red line: fit with $\alpha=1$. Orange zone: region spanned by the fitting function when varying $\alpha$ from 0.5 to 2.}
 \label{fig:fig7}
 \end{figure*}

All the center of mass (CoM) damping rates measured in the BEC-BCS crossover and their respective fitting functions to extract $v_{\rm{c}}$ are shown in  Fig. \ref{fig:fig7}.
As indicated in the main text the fit function is $\gamma_{\rm B} = A \Theta (v-v_c) ((v-v_c)/v_{\rm F})^{\alpha}$, where $\Theta$ is the Heaviside function and $v_{\rm F}$ is the Fermi velocity. The $\chi^{2}$ test reveals that most of our data is consistent with $\alpha=1$ as in \cite{miller2007critical, weimer2014critical}.  Due to the current absence of theoretical prediction for $\alpha$ in a trapped system, we allow $\alpha$ to vary between $0.5$ and $2$, and this induces a systematic correlation between the extracted $\alpha$ and $v_c$ that we include in our error bars on $v_c$ shown in Fig. \ref{fig:fig5}.
The fit results for $\alpha=1$ are displayed in Tab. \ref{tab:t1} along with experimental parameters to produce Fig. \ref{fig:fig5}. The error bars given for $v_{\rm{c}}$ give the span of $v_{\rm{c}}$ when changing $\alpha$ from 0.5 to 2 in the fit function. Note the strong decrease of the damping rate towards the BCS regime.

In addition, $c^{\rm B}_s$ depends on the Bose-Bose scattering length which varies with magnetic field and in particular diverges for a magnetic field of $845.5\,\rm G$ (corresponding to $1/k_{\rm F} a_{\rm F}=-0.13$). We therefore show only  the values of $c^{\rm B}_s$  at the six magnetic field values used in our experiments.

Following \cite{abad2014counter} the Bose-Fermi coupling can be characterized by the quantity
$\Delta=\sqrt{\frac{\partial\mu_{\rm{B}}}{\partial n_{\rm{F}}}\frac{\partial\mu_{\rm{F}}}{\partial n_{\rm{B}}}/\frac{\partial\mu_{\rm{B}}}{\partial n_{\rm{B}}}\frac{\partial\mu_{\rm{F}}}{\partial n_{\rm{F}}}}$. As $\Delta$ increases, the interspecies interactions affect more and more the properties of the system and for $\Delta=1$, the mixture is dynamically unstable at rest and demixes. For measurements presented here we typically have $\Delta\simeq15\%$.


\begin{table*}[h]
\begin{tabular}{|c|c|c|c|c|c|c|}
   \hline
  B (G) & 780 & 800 & 816 & 832 & 860 & 880 \\
  \hline
  $a_{\rm{F}}(a_0)$ & $~6.4 \times 10^3~$ & $~11.3 \times 10^3~$ & $~24.0 \times 10^3~$ & $~\infty~$ & $~-16.5 \times 10^3~$ & $~-10.3 \times 10^3~$ \\
   \hline
  $~1/k_{\rm{F}}a_{\rm{F}}~$ & $~0.68\pm0.07~$ & $~0.39\pm0.01~$ & $~0.18\pm0.02~$ & $~0\pm0.002~$ & $~-0.26\pm0.05~$ & $~-0.42\pm0.03~$ \\
  \hline
   $a_{B}(a_0)$ & $~21.3~$ & $~30.8~$ & $~43.3~$ & $~69.5~$ & $~76.0~$ & $~259~$ \\
  \hline
   $c_{\rm{B}}(10^{-2}v_{\rm{F}})$ & $~9.6\pm1.4~$ & $~9.4\pm0.14~$ & $~11.0\pm1.6~$ & $~11.1\pm1.7~$ & $~11.4\pm1.7~$ & $~15.1\pm2.2~$ \\
  \hline
  $v_{\rm{c}}/v_{\rm{F}}$ & $0.17^{+0.06}_{-0.10}$ & $0.38^{+0.02}_{-0.04}$ & $0.35^{+0.04}_{-0.11}$ & $0.42^{+0.08}_{-0.14}$ & $0.54^{+0.02}_{-0.06}$ & $0.40^{+0.10}_{-0.20}$ \\
  \hline
  $A(s^{-1})$ & $14.8\pm 1.4$& $85 \pm 32$ & $24.6 \pm 4.3$ & $17.3 \pm 3.6$ & $30 \pm 11$ & $2.9 \pm 0.5$ \\
  \hline
  $v_{\rm{c}}/c_{\rm s}^{\rm F}$ & $0.53^{+0.19}_{-0.31}$ & $1.11^{+0.06}_{-0.12}$ & $0.99^{+0.11}_{-0.31}$ & $1.17^{+0.22}_{-0.39}$ & $1.46^{+0.05}_{-0.16}$ & $1.05^{+0.26}_{-0.53}$ \\
   \hline
  \end{tabular}
\caption{ Experimental parameters, sound velocity at the center of the Bose gas in an elongated geometry $c_B=\sqrt{\mu_{\rm{B}}/2m_{\rm{B}}}$,  critical velocity $v_{\rm{c}}/v_{\rm{F}}$, damping rate $A(s^{-1})$, and $v_{\rm{c}}/c_{\rm s}^{\rm F}$ for $\alpha=1$ in the BEC-BCS crossover. The typical number of bosons and fermions are constant in the crossover and are respectively $2.5\pm 0.5\times 10^4$ and $2.5\pm 0.5\times 10^5$.  }
\label{tab:t1}
\end{table*}
\vspace{3cm}
\bigskip
\section{Evolution of the velocities in the trap}

We demonstrate here that for a Bose-Fermi superfluid mixture oscillating in a harmonic trap, the ratio $v/c_{\rm s}^{\rm F}$ is maximum when the centers of two clouds coincide. This can be demonstrated in the general case using the equation of state in the BEC-BCS crossover~\cite{navon2010Ground}, but we will derive it here for the simpler case of a polytropic equation of state.

In the frame of the Fermi cloud, we can describe the trajectory of the BEC by the simple harmonic oscillation
\be
z_{\rm B}(t)=Z_0\cos(\omega_{\rm B} t),
\ee
where we have omitted the slow beating of the amplitude $Z_0$ due to the oscillation-frequency difference between bosons and fermions. The velocity of the BEC is then $v(z)=-Z_0\omega_{\rm B}\sin (\omega_{\rm B}t)$, hence
\be
\left(\frac{v(z)}{v(z=0)}\right)^2=\left(1-\frac{z^2}{Z_0^2}\right)\label{eq:vb},
\ee

For a polytropic equation of state, the local sound velocity in the Fermi cloud is given by \cite{Capuzzi2006Sound}
\begin{eqnarray}
c_{\rm s}^{\rm F}(z)^2&=&\frac{\gamma}{\gamma+1}\frac{\mu_{\rm F}(z)}{m_{\rm F}}\\
c_{\rm s}^{\rm F}(z)^2&=&\frac{\gamma}{\gamma+1}\frac{\mu_{\rm F}(0)}{m_{\rm F}}\left(1-\frac{z^2}{z^2_{\rm TF}}\right),\label{eq:cf}
\end{eqnarray}
where $z_{\rm TF}$ is the Thomas-Fermi radius of the cloud, and the local chemical potential $\mu_{\rm F}(z)$ was obtained using the local density approximation.
Combining equations \ref{eq:vb} and \ref{eq:cf}, we then obtain
\begin{equation}
\frac{v(z)^2}{c_{\rm s}^{\rm F}(z)^2}=\frac{v(z=0)^2}{c_{\rm s}^{\rm F}(z=0)^2}\frac{1-z^2/Z_0^2}{1-z^2/z_{\rm TF}^2},
\end{equation}
which is maximum for $z=0$ when $Z_0\le z_{\rm TF}$.

\bigskip
\section{Lomb-Scargle Algorithm}

\begin{figure}
 \centering
 \includegraphics[width=0.6\columnwidth]{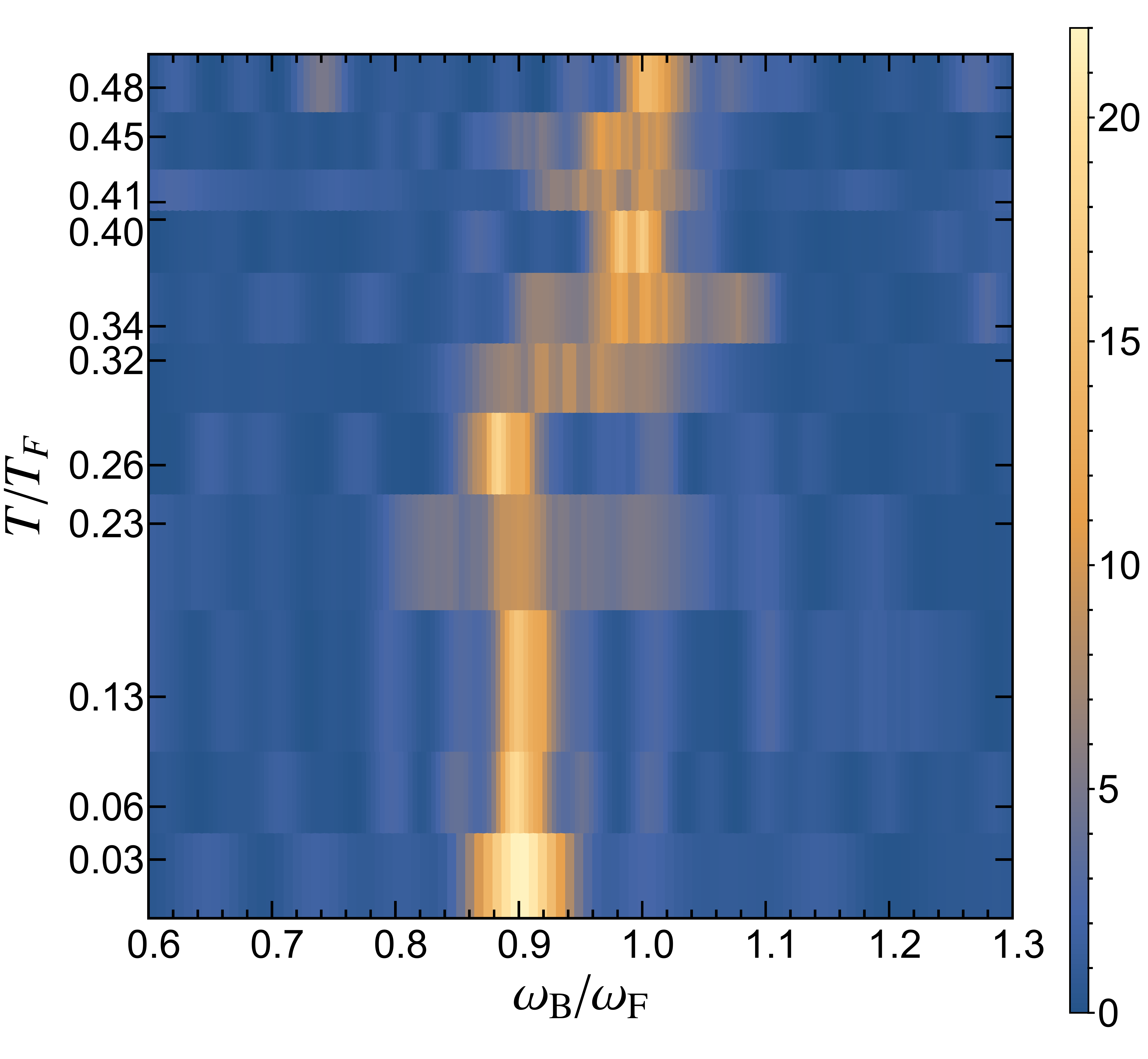}
 \caption{Power spectrum of the oscillations for  different temperatures, obtained using the Lomb-Scargle algorithm of the center-of-mass displacement.  Above $T\approx  T_{\rm c,B}\approx 0.34\,T_{\rm F}>T_{\rm c,F}$, oscillations of the Bose and Fermi clouds become locked together at $\omega_{\rm F}$. A value of 10 for the power represents typically a significance of 0.002.}
   \label{fig:figSpectrum}
 \end{figure}

We use the fit-free Lomb-Scargle periodogram - or Least Square Spectral Analysis - to extract the spectral components of the oscillations for different temperatures \cite{lomb1976fourier,scargle1982fourier}. This method is an adaptation of the Fourier transform to the case of unevenly spaced data.
For $N$ data points $\{h_i=h(t_i)\}_{i=1,\dots, N}$ taken at times $\{t_i\}$, the periodogram is defined as
\begin{equation}
\begin{split}
  P_N(\omega) = \frac{1}{2\sigma^2}\left\{ \frac{[ \sum_j (h_j-\bar{h})\cos\omega (t_j-\tau)]^2}{\sum_j\cos^2\omega(t_j-\tau)}\right. \\
 + \left.\frac{[ \sum_j (h_j-\bar{h})\sin\omega (t_j-\tau)]^2}{\sum_j\sin^2\omega(t_j-\tau)} \right\}
\end{split}
\end{equation}
where $\tau$ is given by $\tan(2\omega\tau)=\frac{\sum_j \sin 2\omega t_j}{\sum_j \cos 2\omega t_j}$ making the periodogram independent of the time origin. $\bar{h}=\frac{1}{N}\sum_{i=1}^N h_i$ and $\sigma=\frac{1}{N-1}\sum_{i=1}^N (h_i-\bar{h})^2$ are the mean and the variance of $\{h_i\}_i$.
The periodogram, or power spectrum (see Fig. \ref{fig:figSpectrum}), gives access to the statistical significance (\textit{ie} the probability of rejecting the null hypothesis when it is true) of each of the evaluated frequencies: noting $P_{\rm max}=\underset{\omega}{\max} ~P_N(\omega)$, the signifiance is proportional to $e^{-P_{\rm max}}$, and here a value of 10 for the power represents typically a significance of 0.002. Fig. \ref{fig:fig2} of the main text displays the set of maxima of Fig. \ref{fig:figSpectrum}; error bars correspond to a significance increased by a factor of 10.

\bigskip

\section{Radiative cascade model}

Consider a mixture of two atomic species labeled by $\alpha=1,2$ of identical masses $m$ and confined in identical harmonic traps. According to Kohn's theorem \cite{kohn1961cyclotron}, the single-species Hamiltonian can always be written as

\be
 H_\alpha = \frac{P_\alpha^2}{2 M_\alpha}+\frac{M_\alpha\omega^2 X_\alpha^2}{2}+H_{\rm int}^{(\alpha)},
\ee
where $P_\alpha$ is the total momentum of cloud $\alpha$, $X_\alpha$ the position of its center of mass and $M_\alpha=N_\alpha m$ its total mass ($N_\alpha$ being the number of atoms). $H_{\rm int}^{(\alpha)}$ acts only on the internal excitation modes of the cloud and commutes with the center-of-mass variables.

Neglecting interspecies interactions, the total Hamiltonian of the system can be written as $H=H_1+H_2$. Introducing the center of mass/relative variables, we have then

\be
H=H_1+H_2=\frac{P^2}{2M}+\frac{M\omega^2 X^2}{2}+\frac{p^2}{2\mu}+\frac{\mu\omega^2 x^2}{2}+H_{\rm int}^{(1)}+H_{\rm int}^{(2)},
\ee
with the usual definitions $P=P_1+P_2$, $X=(M_1 X_1+M_2 X_2)/M$, $p=\mu (P_1/M_1-P_2/M_2)$, $x=x_1-x_2$, $M=M_1+M_2$ and $\mu=M_1M_2/M$. The dynamics of the center of mass and relative variables are described by independent harmonic oscillators decoupled from the internal degrees of freedom. The decoupled base can therefore be written as $|N,n,\varphi\rangle$, where $N$ (resp. $n$) is the excitation number of the center-of-mass (resp. relative) motion, and $\varphi$ describes the state of the internal excitation modes.

Let's now add the interspecies interactions and the mass difference between the two species.

\begin{enumerate}
\item {\em Interspecies coupling}: interactions between the two species are described by the Hamiltonian

\be
H_{1,2}=\sum_{i\le N_1,j\le N_2}U(x_{1,i}-x_{2,j}).
\ee
where $x_{i,\alpha}$ is the position of the $i$-th particle of species $\alpha$ and $U$ is the interspecies interaction potential.

Owing to Kohn's theorem, this Hamiltonian commutes with $P$ and $X$  and therefore  couples  the internal degrees of freedom only to the relative variables $(x,p)$.
\item {\em Mass difference}: assume that species $\alpha$ has a mass $m_\alpha=m+\epsilon_\alpha\delta m/2$, with $\epsilon_1=1$ and $\epsilon_2=-1$. This mass difference adds to the kinetic energy a term

    \be
    \delta H_{\rm K}=-\frac{\delta{m}}{2m}\sum_{i,\alpha}\epsilon_\alpha \frac{p_{i,\alpha}^2}{2m}.
    \ee
As before, we can isolate the center of mass contribution and write
    \be
    \delta H_{\rm K}=-\frac{\delta{m}}{2m}\left[\frac{P^2_1}{2M_1}-\frac{P^2_2}{M_2}\right]+\delta H_{K,\rm int},
    \ee
where $\delta H_{K,\rm int}$ contributes to the internal energies of the clouds and commutes with the center-of-mass degrees of freedom.

\end{enumerate}

Let's now insert these two contributions in the total Hamiltonian. We have

\be
H=H_{\rm CoM}+H_{\rm rel}+H'_{\rm int}+H_{1,2}+H_{\rm coh}.
\ee
with
\begin{eqnarray}
H_{\rm CoM}&=&\frac{P^2}{2M}(1-\rho\frac{\delta m}{2m})+\frac{M\omega^2}{2}X^2\\
H_{\rm rel}&=&\frac{p^2}{2\mu}(1+\rho\frac{\delta m}{2m})+\frac{\mu\omega^2}{2}x^2\\
H'_{\rm int}&=&H_{\rm int}^{(1)}+H_{\rm int}^{(2)}+\delta H_{K,\rm int},\\
H_{\rm coh}&=&\frac{\delta m}{m}\left(\frac{P\cdot p}{M}\right),
\end{eqnarray}
and $\rho=(M_1-M_2)/M$.

This hamiltonian describes two harmonic oscillators ($H_{\rm CoM}$ and $H_{\rm rel}$) coupled to a thermal bath ($H'_{\rm int}$). The coupling is ensured by the $P\cdot p$ terms which couples only the center-of-mass and relative degrees of freedom, and $H_{1,2}$ that commutes with $H_{\rm CoM}$, owing to Kohn's theorem, and couples the relative motion to the internal degrees of freedom of the two clouds.

The interaction between the relative degrees of freedom and the internal thermal bath described by $H_{\rm int}^{1,2}$ leads to an irreversible decay of the relative motion. By contrast,  $H_{\rm coh}$ reflects the coherent beating existing between the relative oscillations of the two species due to their oscillation-frequency difference. It can be expressed using the annihilation operators $a$ and $b$ for the center of mass and relative motions respectively. We have then

\be
H_{\rm coh}=-\frac{\delta m}{m}\hbar\omega\sqrt{\frac{\mu}{M}}\left(a-a^\dagger\right)\left(b-b^\dagger\right).
\ee
Using the rotating wave approximation, one can eliminate the non resonant terms and one finally gets
\be
H_{\rm coh}\simeq\frac{\delta m}{2m}\hbar\omega\sqrt{\frac{\mu}{M}}\left(a^\dagger b+a b^\dagger\right).
\ee
This Hamiltonian is similar to the generalized Caldeira-Leggett model \cite{caldeira1983path} used in solid state physics to study heat transport by phonons in a crystal. The absence of coupling between the bath and Kohn's mode generalizes to the decomposition used in \cite{chou08exact} for a bilinear coupling between the harmonic oscillators and the bath.

In the experiment, we excite the center of mass motion and the initial state is $|N,0,\varphi\rangle$. The coherent coupling transfers the system to the state $|N-1,1,\varphi\rangle$. If the coupling to the bath is strong, the relative motion decays very fast and the system falls into a state $|N-1,0,\varphi'\rangle$ (actually, since several states of the bath are involved, it is more appropriate to describe the state of the two harmonic oscillators by a density matrix rather than a well-defined quantum state). In this case, just like for optical pumping, we can adiabatically eliminate the intermediate state $|N-1,1,\varphi\rangle$ and consider that the dynamics occurs only in the sub-space $|N,0,\varphi\rangle$, where the relative motion is never excited and the centers of mass of the two clouds are locked. In some sense, this freezing of the system state in a pure motion of its center of mass can be considered as a manifestation of the quantum Zeno effect.







\end{document}